\documentclass[
	secnumarabic,
	amssymb,
	nofootinbib,
	nobibnotes,  
	aps,
	12pt, 
	superscriptaddress 
	]
	{revtex4-2}

\usepackage{graphics}
\usepackage{graphicx}
\usepackage{color}
\usepackage{comment}
\usepackage{bm}

\begin{document}

%
%
\title{
Measurement of structure dependent radiative $K^{+} \rightarrow e^{+} \nu \gamma$ decays using stopped positive kaons\\
}

%
%
\author{H.~Ito}
\thanks{
Present address: Department of Physics, Tokyo University of Science, Noda, Chiba 278-8510, Japan  
}
\affiliation{Department of Physics, Chiba University, Chiba, 263-8522, Japan}

\author{A.~Kobayashi}
\affiliation{Department of Physics, Chiba University, Chiba, 263-8522, Japan}

\author{S.~Bianchin}
\affiliation{TRIUMF, Vancouver, BC, V6T 2A3, Canada}

\author{T.~Cao}
\affiliation{Physics Department, Hampton University, VA 23668, USA}

\author{C.~Djalali}
\affiliation
{Department of Physics and Astronomy, University of Iowa, Iowa City, IA 52242, USA} 

\author{D.H.~Dongwi}
\affiliation{Physics Department, Hampton University, VA 23668, USA}

\author{T.~Gautam}
\affiliation{Physics Department, Hampton University, VA 23668, USA}

\author{D.~Gill}
\affiliation{TRIUMF, Vancouver, BC, V6T 2A3, Canada}

\author{M.~D.~Hasinoff}
\affiliation{Department of Physics and Astronomy, University of British Columbia, Vancouver, BC, V6T 1Z1, Canada}

\author{K.~Horie}
\affiliation{Department of Physics, Osaka University, Osaka, 560-0043, Japan}

\author{Y.~Igarashi}
\affiliation{High Energy Accelerator Research Organization (KEK), Tsukuba, 305-0801, Japan}

\author{J.~Imazato}
\affiliation{High Energy Accelerator Research Organization (KEK), Tsukuba, 305-0801, Japan}

\author{N.~Kalantarians}
\thanks{
Present address: Department of Natural Sciences, Virginia Union University, Richmond VA 23220, USA}
\affiliation{Physics Department, Hampton University, VA 23668, USA}

\author{H.~Kawai}
\affiliation{Department of Physics, Chiba University, Chiba, 263-8522, Japan}

\author{S.~Kimura}
\affiliation{Department of Physics, Chiba University, Chiba, 263-8522, Japan}

\author{S.~Kodama}
\affiliation{Department of Physics, Chiba University, Chiba, 263-8522, Japan}

\author{M.~Kohl}
\affiliation{Physics Department, Hampton University, VA 23668, USA}

\author{H.~Lu}
\affiliation{Department of Physics and Astronomy, University of Iowa, Iowa City, IA 52242, USA}

\author{O.~Mineev}
\affiliation{Institute for Nuclear Research, Moscow, 117312, Russia}

\author{P.~Monaghan}
\thanks{
Present address: Department of Physics, Christopher Newport University, Newport News, VA 23606, USA  
}
\affiliation{Physics Department, Hampton University, VA 23668, USA}

\author{S.~Shimizu}
\thanks{
Corresponding author:
suguru@phys.sci.osaka-u.ac.jp}
\affiliation{Department of Physics, Osaka University, Osaka, 560-0043, Japan}

\author{S.~Strauch}
\affiliation{Department of Physics, University of South Carolina, USA}

\author{M.~Tabata}
\affiliation{Department of Physics, Chiba University, Chiba, 263-8522, Japan}

\author{R.~Tanuma}
\thanks{Deceased}
\affiliation{Department of Physics, Rikkyo University, Toshima, 171-8501, Japan}

\author{A.~Toyoda}
\affiliation{High Energy Accelerator Research Organization (KEK), Tsukuba, 305-0801, Japan}

\author{H.~Yamazaki}
\affiliation{High Energy Accelerator Research Organization (KEK), Tsukuba, 305-0801, Japan}

\author{N.~Yershov}
\affiliation{Institute for Nuclear Research, Moscow, 117312, Russia}

\collaboration{J-PARC E36 Collaboration}

%
%
\begin{abstract}
The structure dependent radiative $K^+ \rightarrow e^+ \nu \gamma$ ($K_{e2\gamma}^{\rm SD^+}$) decay was investigated with stopped positive kaons.
The $e^+$ momentum spectra \color{black} containing 574$\pm$30 $K_{e2\gamma}^{\rm SD^+}$ events with \color{black} a $K^+ \rightarrow \mu^+ \nu$ ($K_{\mu2}$)  background of 28$\pm$19 events \color{black} were measured with and without a photon in coincidence and analyzed with Monte Carlo simulations for acceptance and detector response to extract \color{black} the ratio of the branching ratio of the $K_{e2\gamma}^{\rm SD^+}$ decay and  the $K^+ \rightarrow e^+ \nu$  decay including the internal bremsstrahlung process ($K_{e2(\gamma)}$). \color{black}
A value of $Br(K_{e2\gamma}^{\rm SD^+}) / Br(K_{e2(\gamma)}) = 1.12\pm0.07_{\rm stat} \pm 0.04_{\rm syst}$ was obtained.
\color{black}This indicates a partial branching ratio, $Br(K_{e2\gamma}^{\rm SD^+},
~{p_e>200~{\rm MeV}/c},
~{E_\gamma>10~{\rm MeV}})/Br(K_{\mu2})
=(1.85 \pm 0.11_{\rm{stat}} \pm 0.07_{\rm{syst}})
\times10^{-5}$, which is $25\%(\sim$2.5$\sigma)$ higher than the previous experimental result. \color{black}
\end{abstract} 

\maketitle

%
%
\section{Introduction}
High precision measurements of electroweak observables represent powerful tests of the the Standard Model (SM) to obtain hints of new physics~\cite{pdgreview}.
The $K^+ \rightarrow l^+ \nu_l$ ($K_{l2}$) decay, which is one of \color{black}the simplest decays \color{black} among the $K^+$ decay channels, is a clean and sensitive channel to perform such tests. 
Lepton universality signifies identical coupling constants \color{black}for \color{black} the three lepton generations, and \color{black} it \color{black} is a basic assumption in the SM. 
Although each $K_{l2}$ decay width can be described using the $K_{l2}$ hadronic form factor with a few percent accuracy, this form factor can be canceled out by forming the ratio of the \color{black} electronic \color{black} $K^+ \rightarrow e^+ \nu$ ($K_{e2}$) and muonic $K^+\rightarrow \mu^+ \nu $ ($K_{\mu2}$) decay channels ($R_{K}$).

In the $R_K$ determination, the radiative $K^+ \to e^+ \nu \gamma$ decay, which is the $K_{e2}$ decay accompanied with photon emission, has to be taken into account.
There are two $K_{e2\gamma}$ processes~\cite{biji,rev_mod}: the internal bremsstrahlung (IB) process, $K_{e2\gamma}^{\rm IB}$, mostly with low-energy photon emission, and the structure dependent (SD) process, $K_{e2 \gamma}^{\rm SD}$, with high-energy photon emission roughly in the same and opposite directions of the $e^+$ motions, respectively.
In order to compare the experimental value with the SM prediction, the IB process has to be included in the $K_{e2}$ sample (\color{black}$K_{e2(\gamma)}$=$K_{e2}+K_{e2\gamma}^{\rm IB}$ \color{black}) because it is impossible to experimentally separate the IB process from the $K_{e2}$ decay.
The SM prediction, $R_{K}^{ \rm SM}=(2.477\pm0.001)\times 10^{-5}$, can be calculated with excellent accuracy~\cite{RK_theo1, RK_theo2, RK_theo3, RK_theo4}, and this makes it possible to search for new physics effects by a precise $R_{K}$ measurement~\cite{NA62-2013, KLOE2009}.
On the other hand, the SD process, which has a large hadronic uncertainty, is regarded as a background \color{black}for $R_{K}$ \color{black}and has to be subtracted from the observed $e^+$ events. The $K_{e2(\gamma)}$ branching ratio is strongly suppressed down to $\sim$$10^{-5}$ due to the helicity suppression mechanism of the weak charged current. 
The SD process is not subject to the above helicity suppression, and the \color{black}$K_{e2\gamma}^{\rm SD}$ branching ratio \color{black} is comparable to that of $K_{e2(\gamma)}$. 
The SD process is sensitive to the electroweak structure of the kaon and has been the subject of extensive theoretical studies~\cite{biji,rev_mod, chiPT-2004, chiPT-2008}. 

\color{black}In the NA62 experiment~\cite{NA62-2013}, \color{black}which produced the result with the smallest uncertainty, \color{black}in-flight kaon decays in  a \color{black}74\color{black}~GeV/$c$ beam  with a momentum \color{black} spread of $\pm1.4$ GeV/$c$ (rms) \color{black} were used, and the decay particle momentum region from 15 to 65 GeV/$c$ was investigated. \color{black}
The $K_{\mu2}$ and $K_{e2 \gamma}^{\rm SD}$ decays were the main background sources in the $K_{e2(\gamma)}$ sample. 
\color{black}On the other hand, low energy kaons from $\phi \to K^+ K^-$ decays were used in the preceding KLOE experiment~\color{black}\cite{KLOE2009}, \color{black} and the experimental result was dominated by the statistical uncertainty. \color{black}
The NA62 and KLOE results were obtained to be $ R_K=(2.488 \pm 0.007_{\rm{stat}} \pm 0.007_{\rm{sys}})\times 10^{-5}$ and $R_K=(2.493 \pm 0.025_{\rm{stat}} \pm 0.019_{\rm{sys}}) \times 10^{-5}$, respectively, \color{black}and both \color{black} are consistent with the SM prediction within the uncertainties. 
It should be noted that the branching ratio for $K_{e2 \gamma}^{\rm SD}$ reported by the KLOE group was used in the NA62 analysis, and this SD contribution was subtracted from the observed $K_{e2(\gamma)}$ samples. 
Therefore, an experimental check of the SD branching ratio with a systematically different approach from KLOE is important.
In this letter, we present a new measurement of the branching ratio for the $K_{e2 \gamma}^{\rm SD}$ decay relative to that of the $K_{e2(\gamma)}$ decay, Br($K_{e2\gamma}^{\rm SD}$)/Br($K_{e2(\gamma)}$), performed with the J-PARC E36 experiment, which is also aiming at testing lepton universality violation with a precise $R_K$ measurement~\cite{E36-proposal,KAON2013}. 
%
%
 \begin{figure}
  \includegraphics[width=.99\textwidth]
  {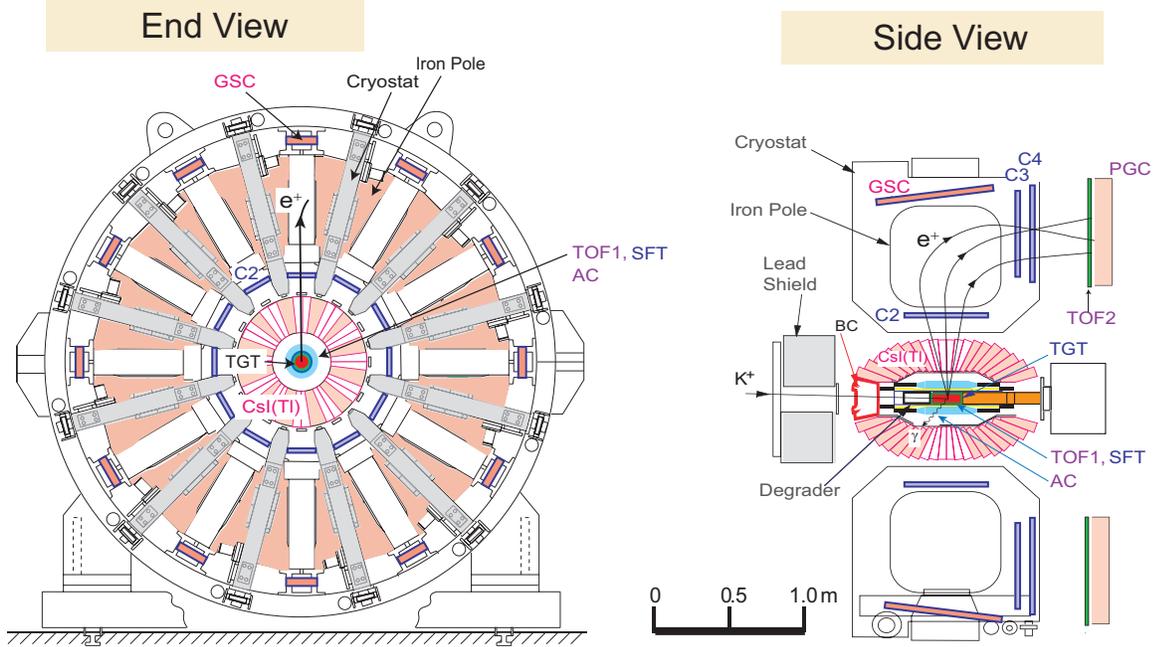}
  \caption{Schematic cross sectional side view (right) and end view (left) of the E36 detector configuration. Charged particles from TGT were \color{black}momentum analyzed \color{black}by reconstructing the particle trajectory using three MWPCs, C2, C3, and C4, as well as by TGT and SFT. Particle identification was carried out using AC, PGC and by measuring the time-of-flight between the TOF1 and TOF2 counters. The photon energy and hit position were measured by the CsI(Tl) calorimeter.}  
  \label{figset}
 \end{figure}

%
%
\section{Experimental details}
\subsection{A stopped $K^+$ beam using the J-PARC K1.1BR beam line}
In contrast to the previous $R_K$ measurements, the E36 experiment used a stopped $K^+$ beam in conjunction with a 12-sector iron-core superconducting toroidal spectrometer~\cite{toro} and a highly segmented CsI(Tl) calorimeter~\cite{E246-CsI(Tl)2000}.
Schematic cross sectional side and end views of the detector configuration are shown in Fig.~\ref{figset}.
Because of the rotational symmetry of the 12 identical gaps in the spectrometer and the large directional acceptance of the $\pi^0$ detector, spectra distortions due to detector acceptance are cancelled and systematic uncertainties are greatly suppressed~\cite{E246-2003}. 
The experimental apparatus was originally constructed for the KEK-PS E246/E470 experiments: a search for $T$-violating muon polarization in $K^+ \rightarrow \pi^0 \mu^+ \nu$ decay~\cite{e246} and spectroscopic studies of various $K^+$ decay channels~\cite{spectros}.

The experiment was performed \color{black}in 2015 \color{black} at the J-PARC Hadron Experimental Facility using a 780~MeV/$c$ separated $K^+$ beam provided by the K1.1BR beam line\color{black}~\cite{doombos}. 
A $K/\pi $ ratio of $\sim$1 was obtained by means of an electrostatic separator system. \color{black}
The $K^+$ beam was discriminated from pion background by a Fitch-type Cherenkov counter (BC)~\cite{Fitch}. 
An efficiency of more than 99\% with a small $\pi$ mis-trigger probability of $<$1\% was achieved for the $K^+$ identification. 
The typical $K^+$ beam intensity was $1.0 \times 10^6$ in a 2-s spill duration and 6-s repetition rate.
\color{black}In total $4.5 \times 10^5$ spills \color{black} were used for the physics production runs. \color{black}
The kaons were slowed down by a degrader and stopped in an active target (TGT), which consisted of 256 3.1$\times$3.1~mm$^2$ thin scintillating bars of 20-cm length forming a cylindrical bundle with a 5.6-cm diameter, located at the center of the detector assembly. 
The $K^+$ stopping efficiency was typically $\sim$0.25 relative to BC $K^+$ triggering, and the $K^+$ stopping profile had a round shape with a Gaussian-like distribution with $\sigma_z$ of $\sim$4 cm in the beam direction. 

%
%
\subsection{Momentum determination of charged particles by the Toroidal spectrometer}
\begin{figure}[hbtp]
\includegraphics[width=\textwidth]{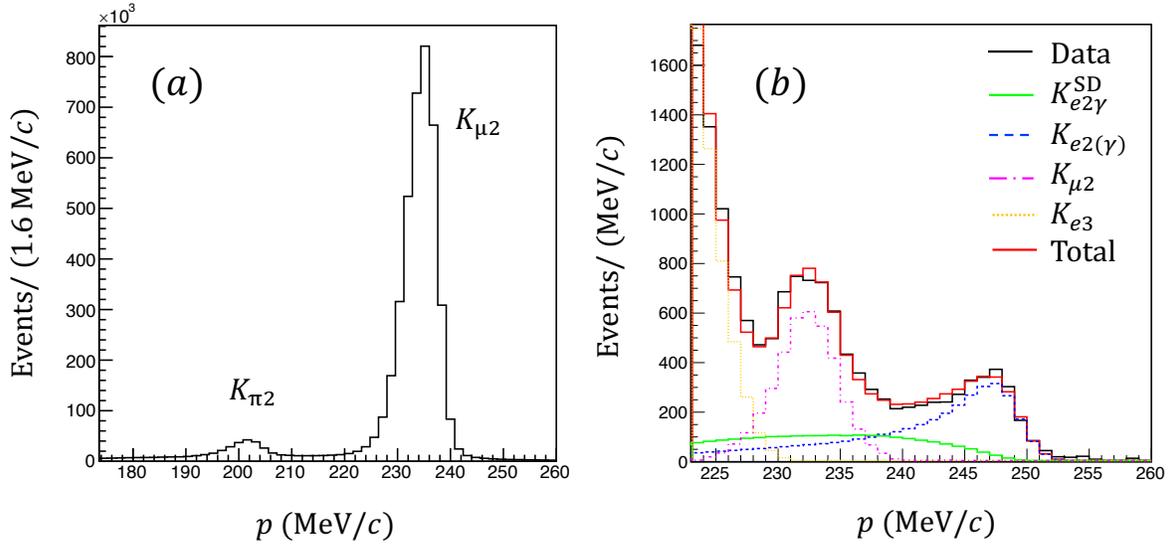}
\caption{
(a) and (b) are the momentum spectra corrected for the energy loss in TGT before and after imposing the positron selection PID, respectively, \color{black}before requiring the photon detection by CsI(Tl). \color{black}
The peak structure due to the predominant $K_{\mu2}$ and $K_{\pi 2}$ decays is seen at 236 MeV/$c$ and 205 MeV/$c$, respectively, in (a).
\color{black}The $K_{\pi2}$ decay is \color{black} reduced due to 
\color{black}
the momentum acceptance of the spectrometer.
\color{black}
The $K_{e2(\gamma)}$ and $K_{e3}$ decays, as well as the remaining $K_{\mu2}$ events due to $\mu^+$ mis-identification are presented in (b). 
The $K_{e2(\gamma)}$ peak is observed with a tail structure in the lower momentum region due to the emission of internal and external bremsstrahlung before entering the spectrometer. 
The momentum in (b) was scaled so that the $K_{e2(\gamma)}$ peak position is at 247 MeV/$c$, and consequently the $K_{\mu2}$ peak position appears at 233 MeV/$c$. 
\color{black}The $e^+$ momentum below 225 MeV/$c$ is \color{black} not usable \color{black} for the $K_{e2\gamma}^{\rm SD}$ and $K_{e2(\gamma)}$ decays due to the \color{black} high \color{black} $K_{e3}$ contribution. \color{black} 
\color{black}The $K_{e2\gamma}^{\rm SD^+}$, $K_{e2(\gamma)}$, $K_{e3}$ and $K_{\mu2}$ decays determined by simulation calculations are also shown in (b). \color{black}
}
\label{fig.mom}
\end{figure}

$K_{e2 \gamma}^{\rm SD}$ and $K_{e2(\gamma)}$ \color{black}candidates \color{black} were identified by analyzing the $e^+$ momentum ($p$) with the 12-sector spectrometer taken under the same trigger and DAQ conditions and, in addition, detecting the photon in the CsI(Tl) calorimeter for $K_{e2 \gamma}^{\rm SD}$. 
\color{black}The trigger condition for event readout was a hit in both TOF counters and in TGT in addition to the $K^+$ beam particle identification by BC. \color{black}
The data were collected at a central magnetic field of $B=1.5$~T, which was optimized for the positron momenta in the region of 220$-$250 MeV/$c$.
Charged particles from TGT were tracked and momentum-analyzed by reconstructing the particle trajectory using multi-wire proportional chambers (MWPCs) located at the entrance (C2) and exit (C3 and C4) of the magnet gap, as well as by TGT and a spiral fiber tracker (SFT) made of scintillating fiber bundles~\cite{E36SFT-2015,tabata_SFT} surrounding TGT. 
The momentum was corrected for the energy loss in TGT assuming that all particles were muons, therefore the $\pi^+$ and $e^+$ momenta after the correction were slightly shifted from their true values. 
The momentum spectrum before imposing the PID analysis is shown in Fig.~\ref{fig.mom}~(a). 
Two peaks due to the $K_{\mu2}$ and $K^+\rightarrow \pi^+ \pi^0$ ($K_{\pi 2}$) decays are clearly visible, \color{black}although the $K_{\pi2}$ events are reduced due to \color{black} the \color{black}lower spectrometer acceptance. \color{black} 
The momentum resolution \color{black}was \color{black}$\sigma_p=2.0$~MeV/$c$ at 236~MeV/$c$. 
The $K^+$ decay time, defined as the \color{black}time of the \color{black}$e^+$ signal at the TOF1 counter, was required to be more than 1.5~ns later than the $K^+$ arrival time determined by BC. 
Small time-of-flight corrections from BC to TGT, and TGT to TOF1 were accounted for on average.
\color{black}The fraction of in-flight $K^+$ decays and any other prompt backgrounds were suppressed down to 0.1\%. \color{black}

%
%

%
%
\subsection{Particle identification}
\begin{figure}[hbtp]
\includegraphics[width=\textwidth]{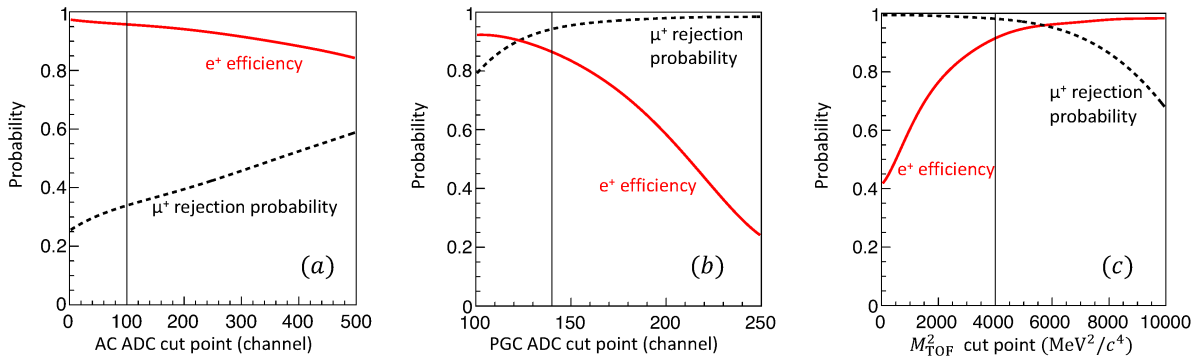}
\caption{
The $e^+$ detection efficiency (solid/red lines) and $\mu^+$ rejection probability (dashed/black lines) \color{black}for $p_{e^+}=247$ MeV/$c$ and $p_{\mu^+}=$ 236 MeV/$c$, \color{black} as functions of the (a) AC, (b) PGC, and (c) $M_{\rm TOF}^2$ cut points. \color{black} The cut points adopted for each detector are also shown\color{black}. As for possible momentum dependence, see the text.
}
\label{fig.pid}
\end{figure}

Particle identification (PID) of $\mu^+$, $\pi^+$, and $e^+$ was carried out in each of the 12 sectors using \color{black}three independent PID systems -- \color{black}an aerogel Cherenkov counter (AC)~\cite{E36AC-2015}, a lead-glass Cherenkov counter (PGC)~\cite{E36PGC-2015}, and by measuring the time-of-flight (TOF) between the TOF1 and TOF2 plastic scintillation counters \color{black}with timing resolutions of 250 ps and 100 ps, respectively. \color{black} 
The AC and TOF1 surrounded TGT while TOF2 was located about 90~cm behind C4 \color{black}resulting in a typical path length of 2.7~m from TOF1. \color{black}
The PGC was placed just after TOF2 at the end of the spectrometer.
Figure~\ref{fig.pid} shows the $e^+$ efficiency (solid/red) and the $\mu^+$ rejection probability (dotted/black) at the momentum of 247~MeV/$c$ and 236~MeV/$c$, respectively, as functions of the (a) AC, (b) PGC, and (c) $M_{\rm TOF}^2$ cut points.
The $e^+$ efficiency \color{black}for each PID system \color{black}was obtained by pre-selecting $e^+$ from the $K^+ \rightarrow \pi^0 e^+ \nu$ ($K_{e3}$) and in-flight $K_{e3}$ decays for the momentum region higher than the $K_{e3}$ endpoint momentum (228~MeV/$c$) by \color{black} using tighter PID conditions than nominally for the other two PID systems. \color{black}
The $\mu^+$ rejection probability was determined using $\mu^+$s from the $K_{\mu2}$ decays. 
Positrons were selected by setting thresholds for AC and PGC at channel 100 and 140, respectively.
Also, the mass-squared of the charged particle ($M_{\rm TOF}^2$) obtained from the TOF, momentum, and path length was required to be $M_{\rm TOF}^2<4000$~(MeV$^2/c^4$).
\color{black}For the AC, PGC, and $M_{\rm TOF}^2$ efficiency determinations, the cut points of (${\rm PGC}>150$, $M_{\rm TOF}^2<3000$~MeV$^2/c^4$), (${\rm AC}>290$, $M_{\rm TOF}^2<3000$~MeV$^2/c^4$), and (${\rm AC}>290$, ${\rm PGC}>150$) were used, and the $\mu^+$ impurity was estimated to be less than 0.5\%.  \color{black}
These positron selection cuts, as shown in Fig.~\ref{fig.pid}, were chosen to remove most of the $K_{\mu 2}$ backgrounds with a $\mu^+$ rejection probability of \color{black}(99.934$\pm0.002_{\rm stat}$)\%, while \color{black} maintaining \color{black} a reasonable $e^+$ efficiency of (75.2$\pm0.4_{\rm stat}$)\%. \color{black}
This was determined to minimize the total uncertainty in the $K_{e2\gamma}^{\rm SD}$ branching ratio measurement from the $K_{\mu2}$ subtraction.
Since the pulse height of the PGC counter increased with increasing $e^+$ momentum and \color{black} the path length depended on the charged particle momentum, this introduced a momentum dependence in the PGC and $M_{\rm TOF}^2$ detection efficiency. On the other hand, the AC efficiency was nearly constant in the observed momentum region.
\color{black}The momentum dependence of the three PID elements \color{black} was \color{black} measured from 200 to 250 MeV/$c$ in order to correct for \color{black} this effect. \color{black}
The black/solid line in Fig~\ref{fig.mom}~(b) shows the charged-particle momentum spectrum with the positron PID condition applied \color{black} and without constraints from the CsI(Tl). \color{black} 
The $K_{e2(\gamma)}$, $K_{e2\gamma}^{\rm SD}$, and $K_{e3}$ decays, as well as the remaining $K_{\mu2}$ events due to $\mu^+$ mis-identification are \color{black}observed, and the momentum \color{black}was slightly scaled so that the $K_{e2(\gamma)}$ peak position is at 247 MeV/$c$. 
The $K_{e2(\gamma)}$ peak has a tail structure in the lower momentum region \color{black}due to the emission of internal and external bremsstrahlung before entering the spectrometer. 
The contribution of $K_{e2(\gamma)}$ events with high energy bremsstrahlung emission was outside of the spectrometer acceptance. 
\color{black} By applying variable cut conditions to suppress $K_{\mu2}$, \color{black}it could be confirmed that there was no $K_{e3}$ tail beyond 230 MeV/$c$. \color{black}

\subsection{Photon measurement by the CsI(Tl) calorimeter}
The photon detector, \color{black}a barrel \color{black}of 768 CsI(Tl) crystals, covered $\sim$70\% of the total solid angle~\cite{E246-2003}. 
There were 12 holes for outgoing charged particles to enter the spectrometer and 2 holes for the beam entrance and exit. Each crystal has a length of 25~cm and covers 7.5$^{\circ}$ in both the polar and azimuthal direction. 
The photon energy and hit position were obtained by summing the energy deposits and calculating the energy-weighted centroid of participating crystals in the Moliere spread.
To read out the CsI(Tl) calorimeter, VF48 Flash ADCs~\cite{fADC} were employed to record the waveform data in order to resolve pulse-pileup events with high efficiency. The hardware threshold was set at $\sim$17~MeV to limit the event size. 
The CsI(Tl) energy and timing resolutions of a single module at 105~MeV were $\sigma_{E}/E \approx 2.6$\% and $\sigma_t= 10.7$~ns~\cite{E36-CsI-2018}, respectively, and the position resolution was obtained as $\sigma_{\rm pos}=7.6$~mm. 
Accidental backgrounds were reduced by choosing a timing window of $\pm 50$ ns.
In addition, some of the photons that passed through the holes in the CsI(Tl) calorimeter into the spectrometer sectors were detected by gap shower counters (GSC), which are sandwich detectors of plastic scintillators and lead plates that will allow us to perform \color{black} a supplemental $K_{e2\gamma}^{\rm SD}$ study by detecting the radiative photons. \color{black}

\section{Analysis}
\subsection{Overview of the \boldmath{$Br(K_{e2 \gamma}^{\rm SD})/Br(K_{e2(\gamma)})$} determination}
In the present study, the SD branching ratio, $Br(K_{e2\gamma}^{\rm SD})$ normalized to that of $K_{e2(\gamma)}$ decay, $Br(K_{e2(\gamma)})$, was determined from the ratio of the $K_{e2\gamma}^{\rm SD}$ and $K_{e2(\gamma)}$ yields, corrected for the detector acceptance as
\color{black}\begin{eqnarray}
\frac{Br(K_{e2\gamma}^{\rm SD})}{Br(K_{e2(\gamma)})}
= \frac{N(K_{e2\gamma}^{\rm SD}) }{ N(K_{e2(\gamma)})} \cdot R_\Omega =
\frac{N(K_{e2\gamma}^{\rm SD}) }{ N(K_{e2(\gamma)})}\cdot
\frac{\Omega(K_{e2(\gamma)})}{\Omega(K_{e2\gamma}^{\rm SD})}, \label{brform}
\end{eqnarray} \color{black}
where $N$ is the number of the accepted events and $R_\Omega$ is the ratio of the overall acceptances $\Omega$ for $K_{e2(\gamma)}$ and $K_{e2\gamma}^{\rm SD}$, respectively, obtained by a Monte Carlo simulation. 
In contrast to the previous KLOE experiment which determined $Br(K_{e2 \gamma}^{\rm SD})$ relative to $Br(K_{\mu 2})$, the present experiment was able to disentangle both the number of $N(K_{e2(\gamma)})$ and $N(K_{e2 \gamma}^{\rm SD})$ events directly from the charged particle momentum spectra. 
The spectrum in Fig.~\ref{fig.mom}~(b) was decomposed by simulating the spectrum of each contributing process and fitting the linear combination to the measured spectrum. 
\color{black}To further constrain $Br(K_{e2\gamma}^{\rm SD})$, separate spectra were obtained for events with 1 and 2 photons detected in the CsI(Tl) calorimeter and for events without conditions on the number of photons; these were fit simultaneously with the ratio $Br(K_{e2\gamma}^{\rm SD}/Br(K_{e2(\gamma)})$ and the yields of $K_{e2\gamma}^{\rm SD}$ and $K_{\mu2}$ decay as free parameters. 
The fit makes use of the $R_{\Omega}$ values from the MC simulation. \color{black}
Our method has the following advantages
: (1) charged particles from the $K_{e2(\gamma)}$ and $K_{e2\gamma}^{\rm SD}$ decays are $e^+$ with \color{black}similar momenta\color{black}, and the PID efficiency up to a small $p$ dependence cancels out; (2) since the $K_{e2(\gamma)}$ decay produces a peak at 247~MeV/$c$ in the momentum spectrum, as shown in Fig.~\ref{fig.mom}~(b), the $K_{e2(\gamma)}$ yield can be \color{black}accurately determined, and, at the same time, the $K_{e2(\gamma)}$ events are largely \color{black}suppressed by requiring a photon hit in the CsI(Tl) calorimeter for the $K_{e2 \gamma}^{\rm SD}$ selection; (3) the CsI(Tl) acceptance can be determined using the two photons from the $K_{\pi2}$ decay; (4) other systematic uncertainties from imperfect reproducibility of the experimental conditions such as tracker inefficiencies, detector misalignment, DAQ deadtime, etc. are also cancelled out in the ratio determination.

\subsection{Detector acceptance}
The detector acceptance for the $K_{e2\gamma}^{\rm SD}$ decays was calculated by a GEANT4-based Monte Carlo simulation assuming the theoretical scheme of vector and axial-vector transitions~\cite{biji,rev_mod}.
The simulation data were generated assuming the Dalitz density, 
\begin{eqnarray}
    \frac{d^2\Gamma(K_{e2\gamma}^{\rm SD})}{dx dy}  
    =  \frac{G_F^2 \alpha  m_K^5 \sin^2\theta_c}{64 \pi^2}
    \times [ (V+A)^2 f_{\rm{SD^+}} + (V-A)^2 f_{\rm{SD^-}} ],
\end{eqnarray}
where $G_F$ is the Fermi constant, $\alpha$ is the fine structure constant, $m_K$ is the kaon mass, and $\theta_c$ is the Cabibbo angle. 
The form factors $V$ and $A$ represent the vector and axial-vector transitions, respectively. The kinematical density distribution for both helicity terms  $f_{\rm{SD^+}}$ and $f_{\rm{SD^-}}$ can be described as
\begin{eqnarray}
    f_{\rm{SD^+}} = (x+y-1)^2(1-x) {~~\rm{and}~~}    &&f_{\rm{SD^-}} = (1-y)^2(1-x),
\end{eqnarray}
by ignoring small ${\cal O}(m_e/m_K)$ contributions, where ${x=2E_\gamma/m_K}$ and ${y=2E_e/m_K}$ are dimensionless photon and $e^+$ energies, respectively, and $m_e$ is the positron mass. 
It should be noted that the SD$^-$, IB, and IB/SD$^+$ interference are negligibly small in the high-momentum $e^+$ region \color{black}$230<{p}<250\rm{~MeV}$/$c$ and in the large ($e^+, \gamma$) opening angle region~\cite{biji}. \color{black}
Here, $V$ was assumed to have the momentum transfer dependence ${V=V_0 [1 +\lambda(1-x)]}$, while $A$ was constant, according to the Chiral Perturbation Theory (ChPT) model at ${\cal O}(p^6)$~\cite{chiPT-2004,chiPT-2008}. 
\color{black}The $\lambda$ and $A/V_0$ parameters were taken to be $\lambda=0.3\pm0.1$ and $A/V_0=0.4\pm0.1$\footnote{\color{black} The difference of the $A/V_0$ value obtained by the ${\cal O}(p^4)$ and ${\cal O}(p^6)$ calculations is adopted as a systematic uncertainty.}\color{black}, respectively, which is the current theoretically conceivable range of ChPT ${\cal O}(p^6)$ model calculations~\cite{rev_mod}.
\color{black}The $K_{e2(\gamma)}$ decay with the IB component, calculated including re-summation of the decay probability for multiple photon emission~\cite{IB_Gatti}, and the $K_{\mu2}$ decay were also generated using the same simulation code. \color{black}

\subsection{Accidental backgrounds in the CsI(Tl) calorimeter}
\begin{figure}[hbtp]
\includegraphics[width=\textwidth]{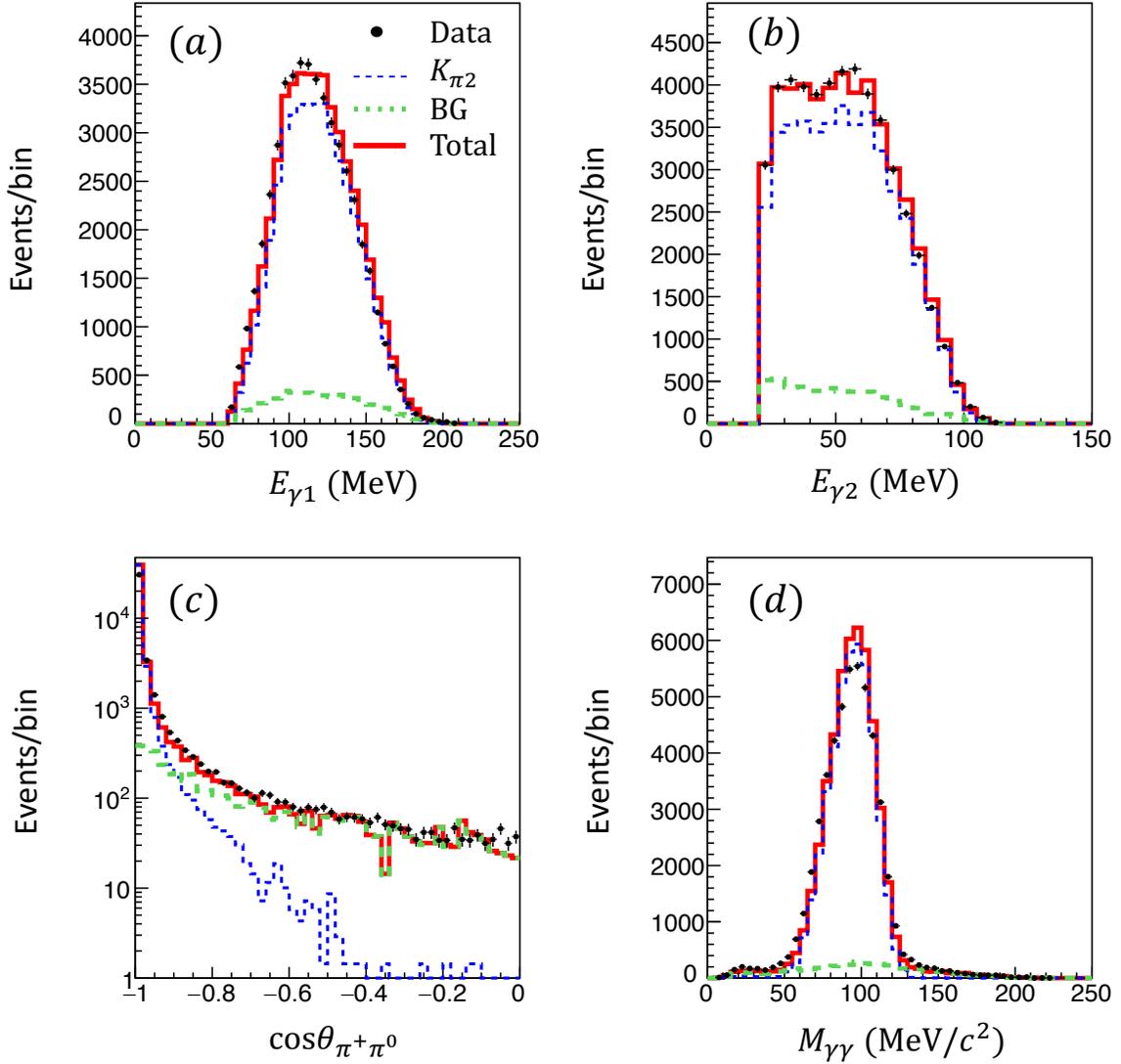}
\caption{
$K_{\pi2}$ spectra compared with the MC simulation taking into account the accidental backgrounds in the CsI(Tl) calorimeter. 
(a) and (b) are the photon energy distributions ($E_{\gamma 1}>E_{\gamma2}$),
(c) is the opening angle between the $\pi^+$ and $\pi^0$,
and (d) is the invariant mass $M_{\gamma\gamma}$.
The black dots are the experimental data. The contribution from the $\pi^0$ decay and events with at least one of the two clusters being accidental are shown as the dotted (blue) and dashed (green) histograms, respectively, and the solid (red) histogram is obtained by summing the two components.
}
\label{fig.kpi2}
\end{figure}

Since the CsI(Tl) calorimeter surrounded the beam axis, it was exposed to a high rate of scattered beam particles and accidental backgrounds in the calorimeter contributed to the raw $K_{e2 \gamma}^{\rm SD^+}$ event samples. 
This accidental background was included in the simulation \color{black} in order \color{black}to reproduce the actual experimental conditions. 
We used the experimental background events and merged them with the simulation data as follows. 
Since the $K_{\mu2}$ decays with $p_{\mu}=236$~MeV/$c$ did not have accompanying photons, the CsI(Tl) signals which coincide with the $K_{\mu2}$ decays within the timing window of $\pm 50$ ns can be treated as pure accidental backgrounds. 
The fraction of the radiative $K^+\to\mu^+\nu\gamma$ decay is negligibly small and causes no effect in this background study. 
The $K_{\mu2}$ events were selected only by the momentum and PID analyses, and these CsI(Tl) signals were merged with the simulation data of the $K_{e2\gamma}^{\rm SD^+}$, $K_{e2(\gamma)}$, and $K_{\mu 2}$ decays. 
It should be noted that the ratio of the single-cluster and zero-cluster $K_{\mu 2}$ events was \color{black}$\epsilon =$(18.85$\pm0.03_{\rm stat}$)\%. \color{black}
The validity of this simulation method was checked using two photons ($E_{\gamma1}>E_{\gamma2}$) from the $\pi^0$ decay in ${K^+\to\pi^+ \pi^0}$ tagged by the $\pi^+$ with $200<p_\pi<210~{\rm MeV}/c$ and the photon energy higher than 21 MeV.
Also, events with large shower leakage from the calorimeter were rejected by requiring $E_{\gamma1}+E_{\gamma 2}>120$ MeV.
Figure~\ref{fig.kpi2} shows the experimental spectra (dots) of (a) $E_{\gamma1}$, (b) $E_{\gamma2}$ , (c) opening angle between the two photons, and (d) invariant mass ($M_{\gamma\gamma}$), together with the simulation data. 
The contribution from the $\pi^0$ decay and events with at least one accidental background hit in the two clusters are shown as the dotted (blue) and dashed (green) histograms, respectively. The solid (red) histogram is obtained by summing the two components and normalizing to the experimental yield.
The results of the simulation are in good agreement with the experimental data, which indicates a good understanding of the photon measurement by the CsI(Tl) calorimeter. 
Also, the detection efficiencies of \color{black}all CsI(Tl) modules \color{black}were determined using the $K_{\pi 2}$ events. Using the information of the $\pi^+$ and one of the two photons, the second photon energy and direction were calculated, and the existence of the actual photon cluster was checked. 

%
%


%
%
\begin{figure}[hbtp]
\includegraphics[width=\textwidth]{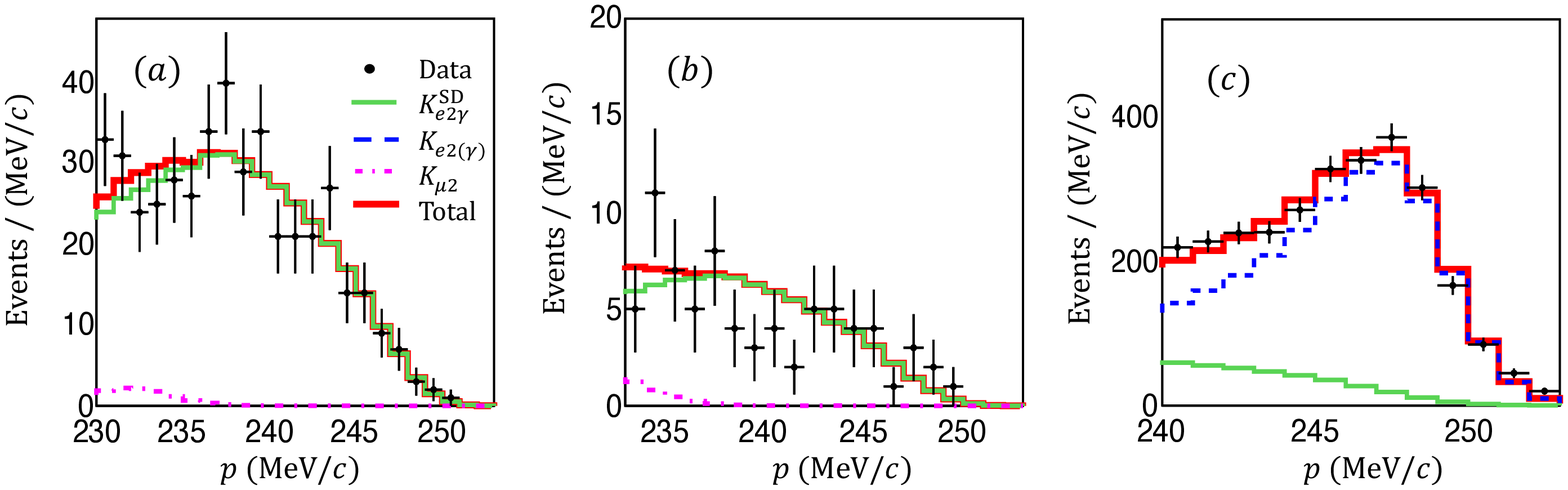}
\caption{
Charged-particle momentum spectra with requiring (a) one photon cluster and (b) two photon clusters in coincidence with the $e^+$ track, and (c) charged particles without any CsI(Tl) constraint.
The dots (black) are the experimental data. The solid (green), dashed (blue), and dashed-dotted (magenta) lines are the $K_{e2\gamma}^{\rm SD^+}$, $K_{e2(\gamma)}$, and $K_{\mu2}$ decays, respectively, determined by simulation calculations. 
The thick-red lines are the fitted results obtained by adding all the decay contributions. 
The events are shown only for the fitted momentum range.
}
\label{fig.nocsi}
\end{figure}

\subsection{\boldmath{$K_{e2\gamma}^{\rm SD^+}$} event selection}
\color{black}In order to relatively enhance the $K_{e2 \gamma}^{\rm SD^+}$ events
and suppress the $K_{\mu2}$ background events, \color{black} photon hits in the CsI(Tl) calorimeter were required.
Due to pile-up of the accidental backgrounds in the CsI(Tl), the accepted $K_{e2 \gamma}^{\rm SD^+}$ events included 2-cluster events in the calorimeter with a ratio of \color{black}the $\epsilon$ probability obtained using $K_{\mu2}$ events \color{black} compared with 1-cluster events.
Since event loss in the 1-cluster data and the appearance of the 2-cluster events were taken into account in the simulation, the $K_{e2 \gamma}^{\rm SD^+}$ branching ratio can be derived by comparing the experimental data with the simulation for both the 1-cluster and 2-cluster events simultaneously. 

%
%
The $K_{e2 \gamma}^{\rm SD^+}$ decays with 1-cluster in the CsI(Tl) were obtained using the following procedure. 
The photon energy and the opening angle between the $e^+$ and $\gamma$ were required to be ${E_{\gamma}>21\;{\rm MeV}}$ and ${\cos\theta_{e\gamma}<-0.8}$.
This $E_{\gamma}$ cut point was a little higher than the hardware threshold to remove effects from small gain variations of each CsI(Tl) module.
Assuming the $K^+ \to e^+ \nu \gamma$ decay kinematics, the missing-mass-squared was calculated as $M^2_{\rm miss}=(m_K - E_e - E_\gamma )^2 - ( {\bf p_e} + {\bf p_\gamma} )^2 $ where ${\bf p}$ is the momentum vector. 
The accepted interval was imposed to be $-4000 < M^2_{\rm miss} < 8000~{\rm MeV}^2/c^4$. 
The momentum spectrum is shown in Fig.~\ref{fig.nocsi}~(a) indicated by the dots. 
Here, a small contribution from $K_{\mu2}$ with an accidental hit remained after the $K_{e2\gamma}^{\rm SD^+}$ selection cuts.
\color{black}On the other hand, the $K_{e2(\gamma)}$ events with an accidental hit were efficiently removed by the $K_{e2\gamma}^{\rm SD^+}$ selection cuts, and the fraction is negligibly small. \color{black}
The decays in the 2-cluster data were selected in a similar manner. 
If one of the two clusters satisfied the conditions for the 1-cluster analysis, the event was adopted as a $K_{e2\gamma}^{\rm SD^+}$ decay and
the associated CsI(Tl) cluster was chosen as the true photon event, as shown in Fig.~\ref{fig.nocsi}~(b). 
It should be noted that the $K_{\mu2}$ surviving fraction relative to the $K_{e2\gamma}^{\rm SD^+}$ yield in the 2-cluster data is approximately twice that observed in the 1-cluster data because there are two photon candidates in the 2-cluster analysis.

%
\subsection{\boldmath{$Br(K_{e2\gamma}^{\rm SD^+})/Br(K_{e2(\gamma)})$} determination}
The $Br(K_{e2\gamma}^{\rm SD^+})/Br(K_{e2(\gamma)})$ value was obtained to be $1.14\pm0.07$ \color{black} for the Prun data set (as defined below; see Table~\ref{branch.result.summary}) \color{black} by simultaneously fitting the momentum spectra of the events with 1-cluster, 2-cluster, and without any CsI(Tl) constraint using the simulation data of the $K_{e2 \gamma}^{\rm SD^+}$, $K_{e2(\gamma)}$, and $K_{\mu2}$ decays, as shown in Fig.~\ref{fig.nocsi}~(a)(b)(c).
\color{black}Here, value of $\epsilon$ obtained with the $K_{\mu2}$ events was used as a constraint in the fit. \color{black}
The solid (green), dotted (blue), and dashed-dotted (magenta) lines are the decomposed $K_{e2 \gamma}^{\rm SD^+}$, $K_{e2(\gamma)}$, and $K_{\mu2}$ events. 
The thick-red line is the fit result obtained by adding all the decay contributions. 
The fitting regions of $p>$ 230, 232, and 240 MeV/$c$ for the events with 1-cluster, 2-cluster, without any CsI(Tl) constraint, respectively, \color{black} were chosen to reduce the effects from the $K_{\mu 2}$ subtraction to minimize the uncertainty of $K_{e2\gamma}^{\rm SD^+}$ \color{black} by eliminating most of the $K_{\mu 2}$ events. 
\color{black}Note that it is very difficult to reproduce these surviving $K_{\mu2}$ events after the PID selection and the $M^2_{\rm miss}$, cos$\theta_{e\gamma}$, $E_{\gamma}$ cuts by the simulation. \color{black}
The $Br(K_{e2\gamma}^{\rm SD^+})/Br(K_{e2(\gamma)})$ result as well as the accepted $K_{e2\gamma}^{\rm SD^+}$ and $K_{e2(\gamma)}$ yields \color{black}used in the fitting \color{black} and the associated $R_\Omega$ values are given in Table~\ref{branch.result.summary} under the heading "Prun" (physics run), along with the statistical uncertainties from the fits.
The statistical uncertainty of $R_\Omega$ obtained from the MC calculation was less than $10^{-3}$. \color{black}

\begin{figure}[hbtp]
\includegraphics[width=\textwidth]{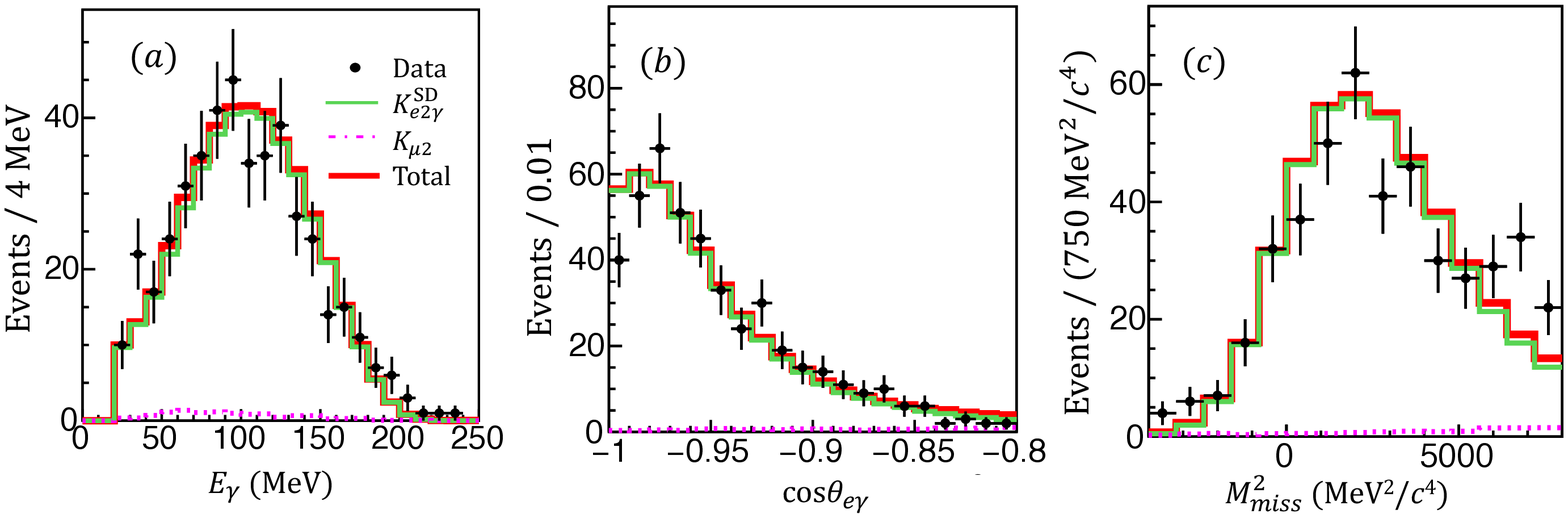}
\caption{
The $K_{e2 \gamma}^{\rm SD^+}$ spectra with the 1-cluster requirement: (a) $E_{\gamma}$, (b) cos$\theta_{e \gamma}$, and (c) $M_{\rm miss}^2$.
The $K_{e2 \gamma}^{\rm SD^+}$ events were selected by imposing $p>230$~MeV/$c$, $-4000<M_{\rm miss}^2<8000~{{\rm MeV^2}/c^4}$, and cos$\theta_{e \gamma}<-0.8$ to suppress the $K_{e3}$ and $K_{\mu2}$ contributions.
The black dots are the experimental data. 
The solid (green) and dashed-dotted (magenta) histograms are the simulation data of $K_{e2 \gamma}^{\rm SD^+}$ and $K_{\mu2}$ with accidental backgrounds, respectively.
The thick-red line is the total simulation result obtained by adding each component. 
}
\label{fig.1gamma}
\end{figure}

The events in Fig.~\ref{fig.nocsi}~(a) were used for an event selection validity check.
Figure~\ref{fig.1gamma} shows the distribution of (a) $E_{\gamma}$, (b) cos$\theta_{e \gamma}$, and (c) $M_{\rm miss}^2$. 
The $K_{\mu2}$ background fraction in Fig.~\ref{fig.1gamma} was successfully suppressed down to $\sim$2\% of the $K_{e2\gamma}^{\rm SD^+}$ yield \color{black}in the fitted momentum range\color{black}.
The experimental data (dots) are in good agreement with the simulation (thick-solid/red), indicating a correct understanding of the $K_{e2 \gamma}^{\rm SD^+}$ acceptance. 
The decomposed $K_{e2 \gamma}^{\rm SD^+}$ (solid/green) and $K_{\mu2}$ (dashed-dotted/magenta) contributions are also shown.

\begin{table}[htbp]
\center
\caption{
Results of the individual counts $N$, acceptance ratio $R_{\Omega}$, and $Br(K_{e2\gamma}^{\rm SD^+})/Br(K_{e2(\gamma)})$ values with statistical uncertainties obtained by simultaneously fitting the events with 1-cluster, 2-cluster, and without any CsI(Tl) constraint for Prun and Crun. 
An error-weighted average of the Prun and Crun results was adopted as the final result. Also, \color{black} $N(K_{e2(\gamma)})$, $N(K_{e2 \gamma}^{\rm SD^+})$, $N(K_{\mu2})$ in the fitting regions of $p>$ 230, 232, and 240 MeV/$c$ for the events with 1-cluster, 2-cluster, without any CsI(Tl) constraint, respectively, are given. \color{black}
}
\begin{tabular}{llccc} 
\hline
&Run period & Prun & Crun & Combined\\
\hline
Without CsI(Tl) constraint&$N(K_{e2(\gamma)})$ & $2353\pm55$ & $330\pm21$ & $2684\pm59$\\
&$N(K_{e2\gamma}^{\rm SD^+})$ & $355\pm19$ & $44\pm7$ & $399\pm20$ \\ 
\hline
%
%
1 cluster
&$N(K_{e2\gamma}^{\rm SD^+})$ & $432\pm24$ & $56\pm9$ & $488\pm26$\\
&$N(K_{\mu 2})$ & $11\pm16$ & $11\pm7$ & $22\pm17$\\
&\color{black}$R_{\Omega_1}$ & 6.22 & 5.83 & \\
\hline
%
%
2 cluster&$N(K_{e2 \gamma}^{\rm SD^+})$ & $77\pm4$ & $9\pm1$ & $86\pm4$\\ 
&$N(K_{\mu 2})$ & $3\pm5$ & $3\pm3$ & $6\pm6$\\
&\color{black}$R_{\Omega_2}$ & 34.8 & 38.4 &\\
\hline

Results&$\chi^2/dof$ & 36.7/43 & 51.7/43 & \\
&$N(K_{e2 \gamma}^{\rm SD^+})$ & $509\pm28$ & $65\pm10$ & $574\pm30$\\ 
&$N(K_{\mu 2})$ & $14\pm17$ & $14\pm8$ & $28\pm19$\\
&$Br(K_{e2\gamma}^{\rm SD^+})/Br(K_{e2(\gamma)})$& $1.14\pm0.07$ & $1.0\pm0.2$ & $1.12\pm0.07$\\
  \hline
  \end{tabular}
  \label{branch.result.summary}
\end{table}

%
%

\begin{figure}[hbtp]
\includegraphics[width=\textwidth]{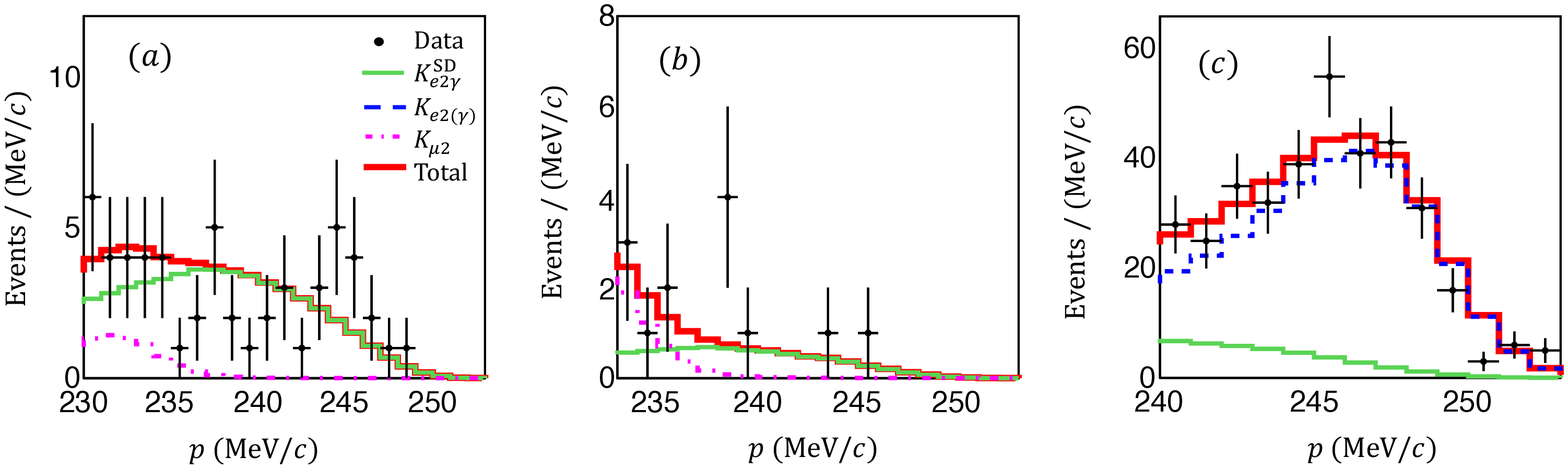}
\caption{
Charged particle momentum spectra for the Crun data 
with requiring the (a) 1-cluster and (b) 2-cluster in the CsI(Tl) calorimeter
and (c) without \color{black} any \color{black} CsI(Tl) constraint. 
The black dots are the experimental data. 
The solid (green), dashed (blue), and dashed-dotted (magenta) lines are the $K_{e2\gamma}^{\rm SD^+}$, $K_{e2(\gamma)}$, and $K_{\mu2}$ decays, respectively, determined by simulation calculations.
The thick-red lines are the fitted result obtained by adding all decay contributions.
}
\label{fig.SCrun}
\end{figure}

%
%
In this experimental study, one of the key issues is the treatment of the accidental background in the CsI(Tl) calorimeter and the $K_{\mu2}$ background that survives after the PID analysis. 
\color{black}In order to validate this analysis method, the data taken during the commissioning runs dedicated to $K^+$ beam and PID detector tuning were used as systematic-control data (Crun), in which the amount of $K_{\mu2}$ background was larger. \color{black}
As a result, the surviving $K_{\mu2}$ fraction in the Crun data was a factor of $\sim$3 higher than in the Prun data.
These data samples were independently analyzed using the same analysis codes adopted for the Prun data. 
The $e^+$ momentum spectra were obtained using the same PID condition for events with the 1-cluster, 2-cluster, and without \color{black} any \color{black} CsI(Tl) constraint, as shown in Fig.~\ref{fig.SCrun}~(a), (b), and (c), respectively, indicated by the dots. 
The $Br(K_{e2\gamma}^{\rm SD^+})/Br(K_{e2(\gamma)})$ ratio was derived to be $1.0\pm 0.2$, which is consistent with the result using the Prun data in spite of the larger number of $K_{\mu 2}$ background events. 
The solid (green), dashed (blue), and dashed-dotted (magenta) lines in Fig.~\ref{fig.SCrun} are the $K_{e2\gamma}^{\rm SD^+}$, $K_{e2(\gamma)}$, and $K_{\mu2}$ decays, respectively, obtained from the simulation. 
The thick-red line is the fit result obtained by adding all the decay contributions. The details of the analysis result are summarized in Table~\ref{branch.result.summary}. 
In addition to the Crun analysis described above, a separate study was performed with the Prun data.
The cuts were tightened to remove \color{black} most of the $K_{\mu2}$ background events \color{black}and relaxed to accept the genuine $K_{e2\gamma}^{\rm SD^+}$ events with higher efficiency, although the statistical uncertainties were significantly enlarged. 
The $Br(K_{e2\gamma}^{\rm SD^+})/Br(K_{e2(\gamma)})$ values determined by these PID conditions were consistent with those obtained using the optimized PID conditions within uncertainties, indicating \color{black} the \color{black} good reproducibility of the PID analysis in the simulation.

%
%

%
%

\begin{table}[htbp]
\center
\caption{Summary of the systematic uncertainties for the
 $Br(K_{e2 \gamma}^{\rm SD^+})/Br(K_{e2(\gamma)})$ ratio determination. }
\begin{tabular}{l r}
\hline
Source & Systematic uncertainty \\ \hline
Hole size of CsI(Tl) calorimeter		& 0.017\\ 
CsI(Tl) misalignment 			        & $<0.001$\\
Imperfect reproducibility of photon angular distribution   	& $<0.001$\\
Accidental backgrounds in CsI(Tl) 		& 0.004\\
Photon energy threshold of CsI(Tl) 		& 0.007\\
Photon energy calibration of CsI(Tl) 	& $<0.001$\\
Photon timing window	            	& 0.009\\
CsI(Tl) detection efficiency            & 0.012\\ 
AC detection(\color{black}PID) \color{black} efficiency  				& 0.007\\ 
PGC detection(\color{black}PID) \color{black} efficiency  				& 0.007\\
TOF detection(\color{black}PID) \color{black} efficiency  				& 0.019\\
$K_{\mu2}$ background subtraction       & 0.015\\
$K_{e2\gamma}^{\rm SD^+}$ form factor 	& 0.011\\
$K^+$ stopping distribution 
                                        & 0.003\\
Material thickness in the central parts
                                        & $<0.001$\\
Positron momentum resolution 	        & 0.002\\
Magnetic field  	        & 0.002\\
In-flight kaon decay 				    & 0.002\\
\hline
Total & 0.036\\
\hline
\end{tabular}
\label{sys.error.summary}
\end{table}

\section{Systematic uncertainties}
In the present work, the $Br(K_{e2 \gamma}^{\rm SD^+})$ value relative to $Br(K_{e2(\gamma)})$ was obtained by calculating the ratio of the $K_{e2\gamma}^{\rm SD^+}$ and $K_{e2(\gamma)}$ yields, as defined in Eq.~\ref{brform}. 
\color{black}The charged particle analysis was \color{black} first \color{black} performed, then the photon measurement was required for the further $K_{e2 \gamma}^{\rm SD^+}$ selection. \color{black}
Therefore, the dominant contributions to the systematic uncertainty are due to the ambiguity of the radiative photon measurement in the $K_{e2 \gamma}^{\rm SD^+}$ decay. 
The systematic uncertainties for the $Br(K_{e2 \gamma}^{\rm SD^+})/Br(K_{e2(\gamma)})$ determination are summarized in Table~\ref{sys.error.summary}. 

The imperfect reproducibility of the CsI(Tl) hole structure \color{black}alignment \color{black}with the 12 spectrometer gaps in the simulation can introduce a systematic uncertainty through a change in the photon acceptance. 
This effect was estimated by considering the maximum conceivable hole size change of 2~mm.
Since the accidental backgrounds were concentrated in the energy region below 30~MeV, these events are very sensitive to the photon energy cut point.
The cut point was changed from 18~MeV to \color{black}40~MeV\color{black}, and the $Br(K_{e2 \gamma}^{\rm SD^+})/Br(K_{e2(\gamma)})$ change was interpreted as the uncertainty from this cut point effect.
Although the CsI(Tl) accidental backgrounds were taken into account in the simulation, the CsI(Tl) timing window was relaxed to accept more accidental events. 
The timing window of $\pm 50$~ns was intentionally increased up to $\pm 60$~ns, and the $Br(K_{e2\gamma}^{\rm SD^+})/Br(K_{e2(\gamma)})$ change was adopted as the systematic uncertainty. 
\color{black}However, if the window was tightened to reduce the accidental backgrounds, the genuine CsI(Tl) events were also rejected and the systematic effect could not be studied. \color{black} 
To check effects from the background intensity fluctuation, the beam background data obtained using the $K_{\mu2}$ events were separated into 4 subsets using time series of the experimental period and \color{black} the \color{black} $Br(K_{e2 \gamma}^{\rm SD^+})/Br(K_{e2(\gamma)})$ was determined for each background sample. 
The variance of the average value was used to estimate this effect.

%
%

The momentum dependence of the PID detectors from 200 to 250~MeV/$c$ was measured using the $K_{e3}$ and in-flight $K_{e3}$ events and taken into account in the simulation. 
However, its statistical uncertainty introduced a possible change in the efficiency correction, which was regarded as a systematic effect in the efficiency correction. 
Also, the statistical uncertainty of the CsI(Tl) efficiency obtained using the $K_{\pi2}$ events was treated as a systematic effect of the photon measurement by the CsI(Tl) calorimeter.
The E36 simulation started from $K^+$ decays at rest and we did not take into account in-flight $K^+$ decay in the simulation, \color{black}because it was not possible to accurately reproduce the $K^+$ stopping process. 
The fraction of the in-flight decays was reduced to less than 0.1\% by the TOF1 timing cut. 
The variation of results observed when this cut was enforced was used as the contribution to the systematic error from this effect. \color{black}
\color{black}In particular, the $K_{e3}$ decays with non-Gaussian tails were carefully checked with and without the $\pi^0$ requirement using the CsI(Tl) calorimeter. \color{black}
Because the $K_{\mu2}$ backgrounds were subtracted from the observed $K_{e2(\gamma)}$ and $K_{e2\gamma}^{\rm SD^+}$ samples, a mis-understanding of the response function for the $K_{\mu 2}$ momentum determination would introduce a systematic uncertainty. 
This effect was estimated by changing the PID conditions around the selected windows.
\color{black}Also, using the MC calculation, \color{black} the effects from \color{black} the $K_{\mu2}$ decays followed by in-flight $\mu^+$ decay were obtained to be negligible\color{black}; they \color{black}can be mainly removed by the tracking information for the momentum determination with additional reduction by the AC and $M_{\rm TOF}^2$ cuts. \color{black}

Since the ($e^+$, $\gamma$) angular correlation and photon energy distributions depend on the $K_{e2 \gamma}^{\rm SD^+}$ form factor, the detector acceptance was affected by the $\lambda$ parameter. 
\color{black}The $Br(K_{e2\gamma}^{\rm SD^+})/Br(K_{e2(\gamma)})$ shift due to a parameter change of $\Delta\lambda=0.1$~\cite{rev_mod} was interpreted as systematic uncertainty. \color{black}
\color{black}Although effects from a $A/V_0$ uncertainty were not serious, the detector acceptance was calculated by varying $\Delta (A/V_0)=0.1$ and treated as a systematic uncertainty. \color{black} 
In addition, effects from a misunderstanding of the $K^+$ stopping distribution and TGT thickness, inaccuracy of the $K^+$ vertex position and $e^+$ momentum determinations, etc. were evaluated, but these effects were common for the $K_{e2(\gamma)}$ and $K_{e2\gamma}^{\rm SD^+}$ decays and cancelled out in their ratio. 
The total size of the systematic uncertainty in the $Br(K_{e2\gamma}^{\rm SD^+})/Br(K_{e2(\gamma)})$ determination was thus obtained by adding each item in quadrature to be 
0.036.

%
%
\section{Result}
The $K_{e2\gamma}^{\rm SD^+}$ branching ratio
relative to the $K_{e2(\gamma)}$ decay was determined using the Prun and Crun analysis results with \color{black}a total of $574\pm 30$ $K_{e2\gamma}^{\rm SD^+}$ events, \color{black}and an error-weighted average of these values was adopted as the final result by adding the total size of the systematic uncertainties, 
$Br(K_{e2 \gamma}^{\rm SD^+})/Br(K_{e2(\gamma)})=1.12\pm0.07_{\rm{stat}}\pm0.04_{\rm{syst}}$.

The $Br(K_{e2 \gamma}^{\rm SD^+})$ value relative to the $K_{\mu2}$ decay can be expressed as 
\begin{eqnarray}
    \frac{Br(K_{e2 \gamma}^{\rm SD^+})}{Br(K_{\mu2})}=\frac{Br(K_{e2 \gamma}^{\rm SD^+})}{Br(K_{e2(\gamma)})} \times
    \frac{Br(K_{e2(\gamma)})}{Br(K_{\mu2})}= \frac{Br(K_{e2 \gamma}^{\rm SD^+})}{Br(K_{e2(\gamma)})} \times R_K^{\rm SM}  
\end{eqnarray}
using the $R_K^{\rm SM}$ prediction. 
Therefore, the $Br(K_{e2 \gamma}^{\rm SD^+})/Br(K_{\mu2})$ value is derived to be $(2.77 \pm 0.17_{\rm{stat}} \pm 0.10_{\rm{syst}}) \times10^{-5}$.
Next, the partial fraction of the $K_{e2 \gamma}^{\rm SD^+}$ branching ratio in the phase space region ($p>200$~MeV/$c$, $E_{\gamma}>10$~MeV) is obtained by correcting for the phase space reduction \color{black}calculated assuming the form factor parameters used in the analysis,
\begin{eqnarray}
\color{black}R_{\gamma}=\color{black}
\frac{Br(K_{e2\gamma}^{\rm SD^+},
~{p>200~{\rm MeV}/c},
~{E_\gamma>10~{\rm MeV}})}{Br(K_{\mu2})}
=(1.85 \pm 0.11_{\rm{stat}} \pm 0.07_{\rm{syst}})
\times10^{-5},
\end{eqnarray}
\color{black}where the systematic effect of this phase space reduction due to the form factor uncertainty is estimated to be 0.007$\times 10^{-5}$ and is already included.
\color{black}This result is almost 25\% ($\sim$2.5$\sigma$) higher than the result\color{black}, $(1.483 \pm 0.066_{\rm{stat}} \pm \color{black}0.013\color{black}_{\rm{syst}})\times 10^{-5}$, reported in the previous experimental study~\cite{KLOE2009} which supported \color{black}the theoretical models of ChPT-${\cal O}(p^4)$, $R_{\gamma}=1.477\times10^{-5}$~\cite{biji}, and ChPT-${\cal O}(p^6)$~\cite{chiPT-2008}. \color{black}On the other hand, the present result is in agreement with the recent lattice calculation, $(1.74 \pm 0.21)\times 10^{-5}$~\cite{lattice_cal}. \color{black}

%
%

%
%
\section*{Acknowledgements}
We would like to express our gratitude to all members of the J-PARC Accelerator, Cryogenic, and Hadron Experimental Facility groups for their support. 
\color{black} A. Kobayashi has made some very major contributions to the data analysis for this experiment and the entire collaboration approves its submission as his doctoral dissertation at Chiba University. 
\color{black}
The present work was supported by JSPS KAKENHI Grant numbers JP26287054, JP15K05113, JP22340059, and JP23654088 in Japan; by NSERC (SAPPJ-2017-00034) and NRC (TRIUMF) in Canada; by Department of Energy (DOE) DE-SC0003884, DE-SC0013941, US National Science Foundation (NSF) PHY-1505615 in the United States; and by Russian Science Foundation Grant No. 14-12-00560 in Russia. 

%
%

\end{document}